\documentclass[aps,pre,onecolumn,amsmath,superscriptaddress]{revtex4}
\usepackage{graphicx}
\usepackage{bm,color,textcomp}
\usepackage{amsmath,amsfonts,amssymb}

\usepackage{subfig}
\usepackage{graphicx}

\usepackage[normalem]{ulem}

\begin{document}

\title{Dynamics of active particles with space-dependent swim velocity}

\author{Lorenzo Caprini} 
\affiliation{Heinrich-Heine-Universit\"at D\"usseldorf, Institut f\"ur Theoretische Physik II - Soft Matter, 
D-40225 D\"usseldorf, Germany.}
\author{Umberto Marini Bettolo Marconi}
\affiliation{School of Sciences and Technology, University of Camerino, Via Madonna delle Carceri, I-62032, Camerino, Italy.}
\author{Ren\'e Wittmann}
\affiliation{Heinrich-Heine-Universit\"at D\"usseldorf, Institut f\"ur Theoretische Physik II - Soft Matter, 
D-40225 D\"usseldorf, Germany.}
\author{Hartmut L\"owen}
\affiliation{Heinrich-Heine-Universit\"at D\"usseldorf, Institut f\"ur Theoretische Physik II - Soft Matter, 
D-40225 D\"usseldorf, Germany.}

\date{\today}

\begin{abstract}
We study the dynamical properties of an active particle subject to a swimming speed explicitly depending on the particle position. The oscillating spatial profile of the swim velocity considered in this paper takes inspiration from experimental studies based on Janus particles whose speed can be modulated by an external source of light. We suggest and apply an appropriate model of an active Ornstein Uhlenbeck particle (AOUP) to the present case.
This allows us to 
predict the stationary properties, by finding the exact solution of the steady-state probability distribution of particle position and velocity. From this, we obtain the spatial density profile and show that its form is consistent with the one found in the framework of other popular models.
The reduced velocity distribution highlights the emergence of non-Gaussianity in our generalized AOUP model which becomes more evident as the spatial dependence of the velocity profile becomes more pronounced.
Then, we focus on the time-dependent properties of the system.
Velocity autocorrelation functions are studied in the steady-state combining numerical and analytical methods derived under suitable approximations. We observe a non-monotonic decay in the temporal shape of the velocity autocorrelation function which depends on the ratio between the persistence length and the spatial period of the swim velocity.
Finally, we numerically and analytically study the mean square displacement and the long-time diffusion coefficient.
The ballistic regime, observed in the small-time region, is deeply affected by the properties of the swim velocity landscape which induces also a crossover to a sub-ballistic but superdiffusive regime for intermediate times.
Finally, 
the long-time diffusion coefficient decreases as the amplitude of the swim velocity oscillations increases because the diffusion is mainly determined by those regions where the particles are slow.
\end{abstract}

\newcommand{\betadef}{\frac{1}{\tau}}
\newcommand{\alphadef}{\frac{\omega_q^2}{\gamma}}
\newcommand{\br}{{\bf r}}
\newcommand{\bu}{{\bf u}}
\newcommand{\bR}{{\bf x}}
\newcommand{\bRz}{{\bf x}^0}
\newcommand{\bk}{{ \bf k}}
\newcommand{\bx}{{ \bf x}}
\newcommand{\vv}{{\bf v}}
\newcommand{\nb}{{\bf n}}
\newcommand{\mb}{{\bf m}}
\newcommand{\bq}{{\bf q}}
\newcommand{\rb}{{\bar r}}

\newcommand{\eeta}{\boldsymbol{\eta}}
\newcommand{\xxi}{\boldsymbol{\xi}}

\maketitle

\section{Introduction}

Nowadays, special active matter systems~\cite{bechinger2016active, marchetti2013hydrodynamics, elgeti2015physics, gompper20202020}, such as engineered E. Coli bacteria and artificial Janus colloids, can be controlled by external stimuli~\cite{walter2007light, buttinoni2012active, dai2016programmable, li2016light, arlt2019dynamics, uspal2019theory}.
For instance, by tuning the power illumination of light, the swim velocity of each active particle can be increased or reduced~\cite{lozano2016phototaxis, gomez2017tuning} and self-assembly such as ``living'' clusters~\cite{palacci2013living} or active molecules~\cite{schmidt2019light} are observed.
An approximate linear relation between light intensity and swim velocity~\cite{lozano2016phototaxis} makes possible a strong control on the parameter of the motility and allows designing complex spatial patterns of the swim velocity landscapes.
This experimental advance offers intriguing perspectives in the world of active matter and provides interesting applications,
ranging from micro-motors~\cite{maggi2015micromotors, vizsnyiczai2017light} and rectification devices~\cite{stenhammar2016light, koumakis2019dynamic} to motility-ratchets~\cite{lozano2019propagating}, where an asymmetric spatial profile of the light intensity is used to induce an asymmetric spatial shape of the swim velocity which produces a net directional motion.
Spatial motility landscapes have been also used to experimentally trap Janus particles~\cite{bregulla2014stochastic, jahanshahi2020realization} and to investigate the occurrence of polarization patterns induced by motility gradients~\cite{soker2021activity, auschra2021density}.
Among the fascinating applications based on light-sensitive active particles, we mention bacteria-based ``painting'', experimentally realized with engineered E. Coli, by Arlt et al.~\cite{arlt2018painting, arlt2019dynamics} and, independently, by Frangipane et al.~\cite{frangipane2018dynamic}, through which some images, such as those of Charles Darwin and Albert Einstein, have been reproduced.
In addition, a numerical study investigates the use of light-sensitive active particles to favor the clustering in a channel geometry and even to induce their clogging, through a sort of plug which can be removed by simply turning off the light~\cite{caprini2020activityclogging}.

From a theoretical perspective, the active Brownian particle (ABP) model has been generalized to the case of non-uniform swim velocity~\cite{sharma2017brownian, ghosh2015pseudochemotactic, liebchen2019optimal, fischer2020quorum} 
also to account for the well-known quorum sensing~\cite{bauerle2018self, jose2021phase, azimi2020bacterial}, chemotaxis and  pseudochemotaxis~\cite{vuijk, merlitz, lapidus, vuijk2021}. This model allows reproducing and predicting one of the leading results concerning the static properties of this system, i.e. a spatial density proportional to the inverse of the swim velocity, originally predicted in the framework of run \& tumble dynamics~\cite{schnitzer1993theory, tailleur2008statistical}.
While it is somehow rather intuitive that a single active particle spends more time in the spatial regions where it moves slowly, in the interacting case, fascinating phenomena can be observed, such as the spontaneous formation of a membrane in two-step motility profiles~\cite{grauer2018spontaneous} and cluster formation in regions with small activity~\cite{magiera2015trapping}.
Many active matter studies even include a temporal dependence in the activity landscape~\cite{maggi2018currents, merlitz2018linear, lozano2019diffusing, geiseler2017taxis} for instance in the form of traveling waves~\cite{zampetaki2019taming, zhu2018transport}. These systems lead to coherent propagation of particle spikes~\cite{lozano2019propagating}, useful to separate binary mixtures~\cite{merlitz2018linear}, and, in some cases, produce counterintuitive directed motion opposite to the propagation of the density wave~\cite{koumakis2019dynamic, geiseler2017self}.

However, while ABP and run \& tumble particles in two dimensions can always be studied numerically, they are quite difficult to be analytically investigated in the presence of a motility landscape: for instance, expressions for the temporal correlations, long-time-diffusion or steady-state dynamical properties remain unknown in those cases.
In analogy with the case of homogeneous swim velocity, an alternative dynamics, known as the active Ornstein-Uhlenbeck particle (AOUP) model, has been proposed by Martin et al~\cite{martin2021statistical} to describe active particles in the presence of spatial-dependent active force. 
In our paper, we propose a new version of the AOUP for space-dependent swim velocity that allows us to obtain more realistic results and further theoretical predictions for static and time-dependent observables.
Our work poses the basis to derive new interesting analytical results in future works, for instance concerning pressure, effective interactions and steady-state probability distributions in the presence of external potentials, which have been analytically studied only in the case of a uniform swim velocity.

The paper is structured as follows: in Sec.~\ref{sec:model}, we introduce the model to describe the dynamics of active particles with space-dependent swim velocity, in particular, the generalization of the Ornstein-Uhlenbeck (AOUP) model. Sec.~\ref{sec:steadystate} and Sec.~\ref{sec:timedependent} report numerical and analytical results in the potential-free case and focus on the steady-state and time-dependent properties, respectively.
In the final section, conclusions are presented.

\section{Model}\label{sec:model}

\subsection{ABP model with spatial-dependent swim velocity}

\begin{figure}[t!]
\includegraphics[width=0.45\columnwidth]{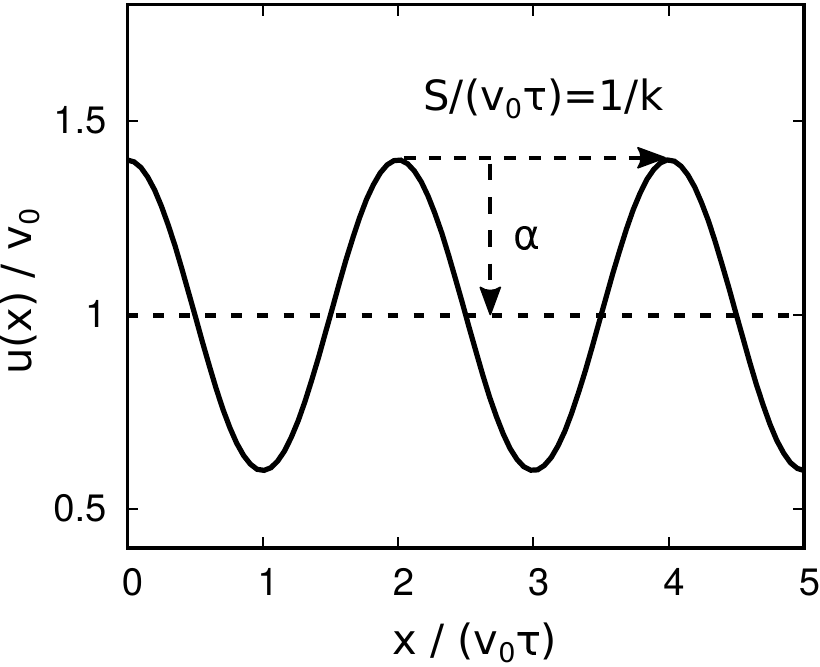}
\caption{Swim velocity profile. The solid black curve shows the spatial profile of the swim velocity $u(x)$, rescaled by the typical particle velocity $v_0$, as a function of the dimension-less position, $x / (v_0 \tau)$, i.e. the position rescaled by the persistence length in the case $u(x)=v_0$.  In the graph, the role of the parameters, $\alpha$ and $k=v_0\tau/ S$ are evidenced. 
}
\label{fig:picture}
\end{figure}

The active Brownian particle (ABP) dynamics represents one of the most established and favorite models to investigate the non-equilibrium behavior of apolar dry active particles~\cite{bechinger2016active, marchetti2013hydrodynamics, shaebani2020computational}. 
It reproduces cluster formation~\cite{mognetti2013living} and motility induced phase separation~\cite{buttinoni2013dynamical, redner2013structure, solon2015pressure, cates2015motility, petrelli2018active, caprini2020spontaneous, shi2020self, turci2021phase, binder2021phase}.
In this model, the active particle is described by a stochastic dynamics for both its position, $\mathbf{x}$, and the degree of freedom called active force, $\mathbf{f}_a$, whose physical origin usually depends on the system under consideration: flagella for bacteria and chemical reactions for Janus particles, to mention just a few examples. 
In the ABP dynamics, $\mathbf{f}_a$ is chosen with constant modulus proportional to the active speed $v_0$ of the particle.
Introducing the friction coefficient, $\gamma$, the active force is:
\begin{equation}
\label{eq:activeforcedefinition}
\mathbf{f}_a=\gamma v_0\hat{\mathbf{n}}  \,,
\end{equation}
where $\hat{\mathbf{n}}$ is a unit vector with components $(\cos\theta, \sin\theta)$ and $\theta$ defines the particle orientation (or direction of the active force) which contains the main stochastic source of the motion, evolving as a Brownian process.
The generalization to a spatial and time dependent swim velocity, can be simply achieved by replacing $v_0\to u(\mathbf{x}, t)$ in Eq.~\eqref{eq:activeforcedefinition}.
Since inertial effects are negligible in most of the experimental systems at the microscopic scale, the dynamics of active particles in two dimensions reads: 
\begin{subequations}
\label{eq:ABP}
\begin{align}
\label{eq:ABP_x}
&\dot{\mathbf{x}}=\sqrt{2 D_t}\mathbf{w}  + u(\mathbf{x}, t)\hat{\mathbf{n}} \,,\\
\label{eq:ABP_theta}
&\dot{\theta} = \sqrt{2 D_r} \xi \,,
\end{align}
\end{subequations}
where $\xi$ and $\mathbf{w}$ are $\delta$-correlated noises with zero average and unit variance, while $D_t$ and $D_r$ are the translational and rotational diffusion coefficients, respectively. 
The inverse of $D_r$ defines the persistence time of the single-particle trajectory, i.e. the average time that the particle spends without changing direction, $\tau = 1/((d-1) D_r)$.
The function $u(\mathbf{x}, t)$ is the swim velocity induced by the active force that, here, is a generic function of both position and time.
The shape of $u(\mathbf{x}, t)$ cannot be chosen arbitrarily but must satisfy the following properties dictated by physical arguments:
\begin{itemize}
\item[i)] $u(\mathbf{x}, t)\geq0$, for every $x$ and $t$, because the swim velocity is positive by definition and is nothing but the modulus of the velocity induced by the active force.
\item[ii)] $u(\mathbf{x}, t)$ needs to be chosen as a bounded function of its arguments because the swim velocity cannot reach an infinite value. 
\end{itemize}
In this work, we restrict to a static profile, so that $u(\mathbf{x}, t) = u(\mathbf{x})$.
While most of the theoretical results are valid for arbitrary $u(\mathbf{x})$, we further consider a one-dimensional profile of the swim velocity to test our predictions by numerical simulations.
Specifically, inspired by the experiments with Janus particles studied in Ref.~\cite{lozano2016phototaxis} and taking in mind properties i) and ii), we choose $u(\mathbf{x})$ as a periodic function of its argument varying along with one spatial coordinate, namely $x$: 
\begin{equation}
\label{eq:swimvelocityprofile}
u(x)=v_0\left(1+\alpha \cos{\left( 2\pi \frac{x}{S}\right)} \right) \,,
\end{equation}
 with the spatially averaged swim velocity $v_0>0$ and the amplitude $\alpha<1$ so that i) and ii) are always satisfied.
The spatial profile of $u(x)$ is shown in Fig.~\ref{fig:picture}. 
The parameter $S>0$ determines the spatial period of the swim velocity, while $v_0(1-\alpha)$ and $v_0(1+\alpha)$ are the minimal and maximal swim velocity, respectively.

\subsection{AOUP model with spatial-dependent swim velocity}

The active Ornstein-Uhlenbeck particle (AOUP) model~\cite{caprini2019activity, marconi2016velocity, wittmann2017effective, dabelow2019irreversibility, berthier2019glassy, woillez2020nonlocal, martin2020statistical} has been introduced in the field of active matter to simplify the ABP dynamics~\cite{fily2012athermal, farage2015effective} but also to describe the behavior of colloidal particles immersed in active baths, for instance formed by bacteria~\cite{maggi2017memory}.
The AOUP model reproduces the typical phenomenology of active particles because a single-particle trajectory shows a certain degree of persistence, the mean square displacement of the two models coincides~\cite{sevilla2015smoluchowski, caprini2020inertial, nguyen2021active}. Both AOUP and ABP active particles accumulate near obstacles~\cite{caprini2018active, das2018confined, caprini2019activechiral} and display the non-equilibrium clustering, known as motility induced phase separation (MIPS)~\cite{fodor2016far, maggi2021universality}.
The AOUP approximation has been employed to obtain a plethora of analytical results which cannot be achieved by using the ABP:
perturbative expressions for the probability distribution of particle position and velocity~\cite{fodor2016far}, approximated formula for the pressure~\cite{marconi2016pressure,wittmann2019pressure}, the effective temperature in confined systems~\cite{szamel2014self}, and the analytical shape of the spatial and temporal velocity correlation function of dense active systems~\cite{caprini2020hidden, szamel2021long, caprini2020time}.

In the AOUP description, the active force is simply described by a two-dimensional Ornstein-Uhlenbeck process and can be generalized
to the case of spatio-temporal swim velocities, as follows:
\begin{subequations}
\label{eq:AOUP}
\begin{align}
\label{eq:AOUP_x}
&\dot{\mathbf{x}}=\sqrt{2 D_t}\mathbf{w}  + u(\mathbf{x}, t){\boldsymbol{\eta}} \,, \\
\label{eq:AOUP_eta}
&\tau\dot{\boldsymbol{\eta}} = - \boldsymbol{\eta} + \sqrt{2\tau}{\boldsymbol{\chi}}  \,.
\end{align}
\end{subequations}
The main simplification with respect to the ABP dynamics~\eqref{eq:ABP} is obtained by replacing the unit vector $\hat{\mathbf{n}}$ with an Ornstein-Uhlenbeck process, $\boldsymbol{\eta}$, with unit variance and typical time $\tau$, which intrinsically defines the persistent time of the trajectory.
The term $\boldsymbol{\chi}$ is a vector of white noises whose components are $\delta$-correlated.
Taking $u(\mathbf{x}, t)=v_0$, we recover the standard AOUP model by assuming $v_0^2=d D_a/\tau$, where $D_a$ is the diffusion coefficient due to the active force and $d$ is the dimension of the system.
Recall that the full connection with the ABP model could be established by fixing $(d-1)\tau=1/D_r$ so that the time-correlation of the AOUP active force coincides with that of the ABP one~\cite{farage2015effective, caprini2019comparative}. 
We also note that the AOUP model 
allows us to consider a one-dimensional system with a spatial velocity profile (as Eq.~\eqref{eq:swimvelocityprofile}) at variance with the ABP model.
In appendix~\ref{app:AOUPMartin}, we discuss the difference between the present model and the alternative AOUP dynamics with space-dependent motility landscape introduced by Martin et al.~\cite{martin2020statistical}.

\subsubsection{Velocity description of AOUP}

Many theoretical advances in the study of the AOUP model stem from the introduction of an auxiliary description of the active particle in terms of position, $\mathbf{x}$, and particle velocities, $\mathbf{v}$,
which can be obtained by performing a change of variable $(\mathbf{x}, \boldsymbol{\eta}) \to (\mathbf{x}, \dot{\mathbf{x}}=\mathbf{v})$.
This simple strategy makes the AOUP model particularly appealing because its
dynamics becomes similar to the well studied dynamics of passive underdamped Brownian particles. 
In the following, we generalize this method to account for the
spatial-dependent swim velocity $u(\mathbf{x}, t)$.
Here, for simplicity, we neglect the thermal noise since the translational diffusivity is usually some order of magnitudes smaller than the active one~\cite{bechinger2016active}, $D_t \ll D_a$.
However, we remind that a generalization of the method to include also the thermal noise can be obtained by following Ref.~\cite{caprini2018active}.
Taking the time-derivative of Eq.~\eqref{eq:AOUP_x}, using Eq.~\eqref{eq:AOUP_eta} to eliminate $\dot{\boldsymbol{\eta}}$ and
$\boldsymbol{\eta}$ 
in favor of $\mathbf{x}$ and $\mathbf{v}$ (using again Eq.~\eqref{eq:AOUP_x}), we get:
\begin{subequations}
\label{eq:dynamics_xv}
\begin{align}
\label{eq:dynamics_x}
\dot{\mathbf{x}}=&\mathbf{v} \,,\\
\label{eq:dynamics_v}
\tau\dot{\mathbf{v}}=& -  \mathbf{v} + u(\mathbf{x}, t) \sqrt{2\tau} \boldsymbol{\chi}\\
&+\tau\frac{ \mathbf{v}  }{ u(\mathbf{x}, t)} \left( \frac{\partial}{\partial t} + \mathbf{v} \cdot \nabla \right) u(\mathbf{x}, t)\,. \nonumber
\end{align}
\end{subequations}
The first line of Eq.~\eqref{eq:dynamics_v} resembles the dynamics in the case of a constant swim velocity
$u(\mathbf{x}, t)=v_0$: the dynamics of an overdamped active particle is mapped onto that of an underdamped passive particle with mass $\gamma \tau$ subject to an inertial force, an effective friction whose amplitude is $1/\tau$ and a stochastic white noise.
However, in the present case, the noise amplitude in Eq.~\eqref{eq:dynamics_xv} contains a spatial and temporal dependence via the function $u(\mathbf{x}, t)$.
The second line provides 
an additional force term accounting both for the time and space-dependence of $u(\mathbf{x}, t)$. 
By choosing a static profile, such that $u(\mathbf{x}, t)=u(\mathbf{x})$, the new term quadratically depends on $\mathbf{v}$ and is spatially modulated by the function $\nabla_x \log(u(\mathbf{x}))$. 
As a consequence, the latter term changes its sign depending on the region of space where the particle is placed, a feature that guarantees the convergence of the dynamics.
We remark that this term cannot be interpreted as a friction force, because of its even parity under time-reversal transformation.

To switch back from the coordinates $(\mathbf{x}, \mathbf{v})$ to $(\mathbf{x}, \boldsymbol{\eta})$, one has to account for the Jacobian of the transformation which, now, is not trivial at variance with the case $u(\mathbf{x}, t)=v_0$.
Indeed, the relation:
\begin{equation}
{\mathbf{v}}=\dot{\mathbf{x}}= u(\mathbf{x}, t) \boldsymbol{\eta}
\end{equation}
implies that the Jacobian is a function of $\mathbf{x}$ and $t$ which reads:
\begin{equation}
|J|=u(\mathbf{x}, t)
\end{equation}
so that the probability distribution in the original coordinates, namely $\tilde{p}(\mathbf{x},\boldsymbol{\eta})$, and the one obtained after the change of variables, namely $p(\mathbf{x}, \mathbf{v})$, satisfy:
\begin{equation}
\tilde{p}(\mathbf{x} ,\boldsymbol{\eta}, t) = u(\mathbf{x}, t)  p(\mathbf{x}, \mathbf{v}, t)\,.
\end{equation}

Finally, we remark that in the case $D_t=0$, $\tau$ provides the natural time-scale to evaluate the time $t$, while $v_0$ determines the one for the particle velocity. With this choice, it is straightforward to recognize that the particle position can be rescaled by the persistence length $v_0 \tau$ of the case $u(\mathbf{x})=v_0$.
We conclude that the dynamics is  controlled by the parameters affecting the value of $u(\mathbf{x})$.
Therefore, choosing the profile of $u(x)$ as in Eq.\eqref{eq:swimvelocityprofile}, we can recognize $S$ as the additional spatial-scale to compare with $v_0 \tau$ (or equivalently $S/v_0$ as the typical time to compare with  $\tau$). Finally, we expect that ratio between the amplitude of the $u(x)$ oscillations and the typical velocity, namely the parameter $\alpha$, will play a crucial role. 
As a consequence, in the case $D_t=0$, the dynamics is affected by two dimensionless parameters, $k=v_0\tau/S$ and $\alpha$.

\section{Steady-state properties}\label{sec:steadystate}

\begin{figure}[t!]
\includegraphics[width=0.45\columnwidth]{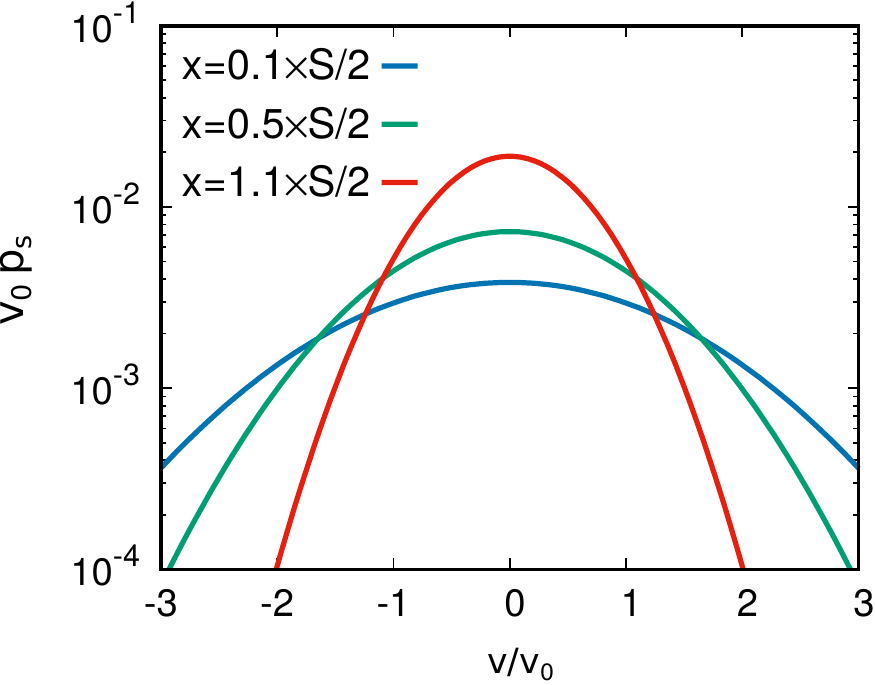}
\caption{Steady-state probability distribution function, $p_s(x,v)$, as a function of the rescaled velocity, $v/v_0$, calcutated at different positions as shown in the legend. Points are obtained from numerical simulations while solid lines by plotting the prediction~\eqref{eq:potentialfree_solution}. 
The curves are obtained by considering a box of length $L/(v_0 \tau)=10$ and by setting $\alpha=2/5$ (results are independent of $k=5/2$).
}
\label{pxv_final.eps}
\end{figure}

Hereafter, we restrict to the static case $u(\mathbf{x}, t)=u(\mathbf{x})$.
From the stochastic dynamics~\eqref{eq:dynamics_xv}, it is straightforward to derive the Fokker-Planck equation for the probability distribution  $p=p(\mathbf{x}, \mathbf{v}, t)$:
\begin{flalign}
\label{eq:FokkerPlanck_potentialfree}
\partial_t p = & \nabla_v \cdot \left[ p \frac{\mathbf{v}}{\tau}   \right] + \frac{u^2(\mathbf{x}) }{\tau} \nabla^2_{v^2}   p\\
&-\mathbf{v} \cdot \nabla_x p - \frac{1}{u(\mathbf{x})}\nabla_v \cdot \left[ p\,\mathbf{v} \left(\nabla_x u(\mathbf{x}) \cdot \mathbf{v}\right)  \right]\,,\nonumber 
\end{flalign}
where $\nabla_x$ and $\nabla_v$ are the spatial and velocity gradient, respectively.
In the steady-state, Eq.~\eqref{eq:FokkerPlanck_potentialfree} admits a local Gaussian solution of the form:
\begin{equation}
\label{eq:potentialfree_solution}
p_s(\mathbf{x},\mathbf{v}) = \frac{\mathcal{N}}{\sqrt{2\pi} u^2(\mathbf{x})} \exp{\left( - \frac{\mathbf{v}^2}{2u^2(\mathbf{x})}  \right)} \,,
\end{equation}
where $\mathcal{N}$ is a normalization constant. 
This result directly generalizes the Gaussian solution obtained for a uniform swim velocity~\cite{marconi2016velocity, fodor2016far, szamel2014self}.
Considering the system on a box of size $L^d$ with periodic boundary conditions, $\mathcal{N}$ reads:
\begin{equation}
\label{eq:normalizationN}
\mathcal{N}^{-1}=\int_{L^d}\,\frac{d\mathbf{x}}{u(\mathbf{x})}  \,.
\end{equation}
Specifically, for the periodic one-dimensional profile of $u(x)$, Eq.~\eqref{eq:swimvelocityprofile}, the integral in Eq.~\eqref{eq:normalizationN} can be analytically calculated and reads $\mathcal{N}^{-1}=L/v_0/\sqrt{1-\alpha^2}$, and is independent of $S$.
In this case, the stationary mean-squared velocity, $\langle v^2(x)\rangle$, depends explicitly on $x$ through the profile of $u(x)$ and, therefore, has an oscillatory shape.
Figure~\ref{pxv_final.eps} shows $p_s(v,x)$ as a function of $v$ for different values of $x$. 
The distribution $p_s(v,x)$ becomes narrower (smaller variance) as far as $x$ is increased until it starts to decrease again for $x>S/2$
in agreement with the periodicity of $u^2(x)$.

\subsection{Spatial density and hydrodynamic approach}

\begin{figure*}[t!]
\includegraphics[width=0.96\textwidth]{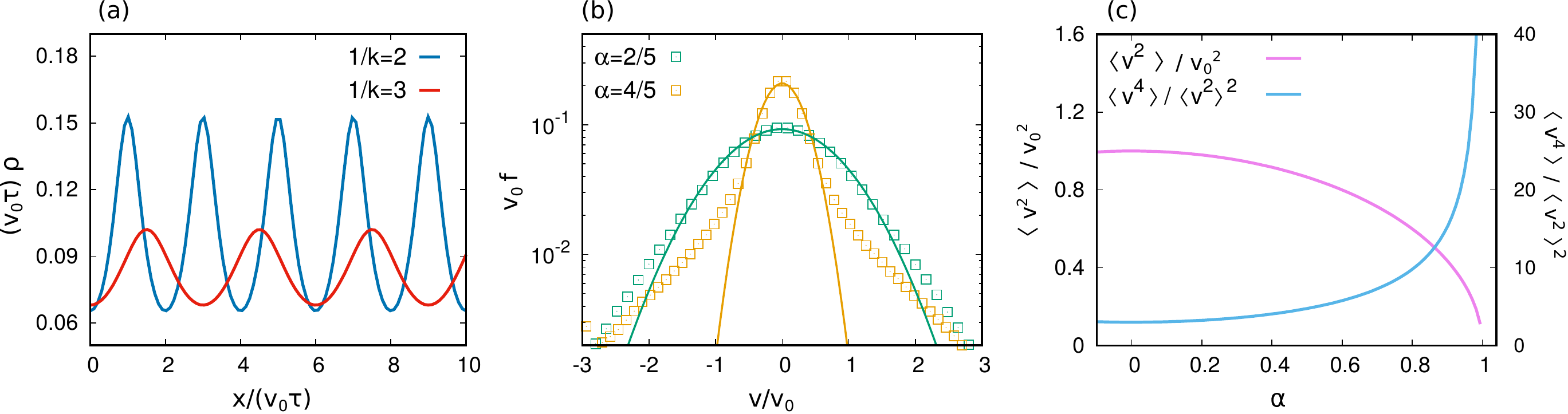}
\caption{
Reduced probability distributions. Panel~(a): density, $\rho(x)$, as a function of $x/(v_0 \tau)$ (position over persistence length), for two different profiles of the swim velocity $u(x)$, i.e. varying $k=\tau v_0 / S$. 
Solid lines from the exact expression~\eqref{eq:density}  with $\alpha=2/5$.
Panel~(b): marginal velocity distribution, $f(v)$, as a function of $v/v_0$, for two different values of $\alpha$, i.e. varying the amplitude of the oscillatory part of $v_0(x)$. In this case, the results does not depend on the value of $k$.  
Points are obtained from numerical simulations while solid lines by fitting a Gaussian function of the form $\sim a \exp(-x^2/b)$ where $a$ and $b$ are two fitting parameters (mind the logarithmic scale on the vertical axis).
%
%
Panel~(c) plots the variance and the kurtosis of the reduced velocity distribution, $\langle v^2\rangle$ and $\langle v^4\rangle/\langle v^2\rangle^2$, respectively, as a function of $\alpha$ (also in this case, $k$ does not play any role). Solid lines are obtained from Eqs.~\eqref{eq:variance_v} and~\eqref{eq:kurtosis_v} for the variance and kurtosis, respectively. Black dashed lines are guides for eyes.
Finally, simulations are obtained considering a box of length $L/(v_0 \tau)=10$ with periodic boundary conditions. }
\label{fig:marginal_prob_final}
\end{figure*}

By integrating out the velocity in the solution \eqref{eq:potentialfree_solution}, one can straightforwardly derive the steady-state density function, $\rho(x)$, obtaining the profile: 
\begin{equation}
\label{eq:density}
\rho(\mathbf{x}) = \int d\mathbf{v} p_s(\mathbf{v},\mathbf{x}) = \frac{\mathcal{N}}{u(\mathbf{x})} \,,
\end{equation}
where the normalization constant $\mathcal{N}$ is given by Eq.~\eqref{eq:normalizationN} for a simulation in a box of size, $L^d$, with periodic boundary conditions.
%
This exact result is shown in Fig.~\ref{fig:marginal_prob_final}~(a), where we plot $\rho(x)$, i.e. the density distribution along $x$,
by chosing the spatial profile of $u(\mathbf{x})$ as in Eq.~\eqref{eq:swimvelocityprofile}, i.e. varying only along one coordinate.
The density $\rho(x)$ is reported for two different values of $k$ which determines the period of its oscillation.

To the best of our knowledge, Eq.~\eqref{eq:density} is consistent with the only analytical result obtained for an active particle with spatial-dependent swim velocity. 
By resorting to a hydrodynamic approach, the spatial profile of $\rho(\mathbf{x})$ has been derived first for run \& tumble dynamics~\cite{schnitzer1993theory, tailleur2008statistical} and, successively, for active Brownian particles. 
%
To develop the hydrodynamics of our generalized AOUP, we adopt a strategy similar to that used in Ref.~\cite{tailleur2008statistical}. 
We start by introducing the first two conditional moments (i.e. at a fixed position) of the velocity distribution, namely the momentum density field:
\begin{equation}
\mathbf{m}(\mathbf{x}, t) 
= \frac{1}{\rho(\mathbf{x},t)}\int d\mathbf{v}\, p(\mathbf{v},\mathbf{x}, t)\, \mathbf{v}  \,,
\end{equation}
and the velocity tensor:
\begin{equation}
\mathbf{Q}(\mathbf{x}, t)
 = \frac{1}{\rho(\mathbf{x}, t)}\int dv \,p(\mathbf{v}, \mathbf{x}, t) \,\mathbf{v}\mathbf{v} \,.
\end{equation}
By integrating out the velocity in Eq.~\eqref{eq:FokkerPlanck_potentialfree}, we obtain the continuity equation for the density field which, after dropping the explicit dependence on $\mathbf{x}$ and $t$, reads:
\begin{equation}
\label{eq:continuityeq}
\partial_t \rho=- \nabla_x ( \mathbf{m}\, \rho) \,,
\end{equation}
while, multiplying Eq.~\eqref{eq:FokkerPlanck_potentialfree} by $\mathbf{v}$ and, then, integrating on $\mathbf{v}$, we obtain the momentum balance equation:
\begin{equation}
\label{eq:momentumbalance}
\begin{aligned}
&\partial_t \left[\mathbf{m} \,\rho\right]= -\frac{\mathbf{m}\, \rho}{\tau}  - \nabla_x \cdot \left(\mathbf{Q} \,\rho\right) + \frac{\rho}{u} \, \mathbf{Q} \cdot \nabla_x u  \,.
\end{aligned}
\end{equation}
The steady-state solution admitted by Eqs.~\eqref{eq:continuityeq} and~\eqref{eq:momentumbalance} is consistent with the shape of $\rho(\mathbf{x})$, given by Eq.~\eqref{eq:density}, upon assuming a Gaussian closure of the moments hierarchy and, specifically, upon recognizing that, in the steady-state, $ \mathbf{m}=0$ and $\mathbf{Q}=\mathbf{I} \,u^2$, when $\mathbf{I}$ is the identity matrix.
Of course, the Gaussian closure is perfectly suitable to exactly close the hydrodynamics hierarchy because of the Gaussian shape of Eq.~\eqref{eq:potentialfree_solution}.

\subsection{Reduced velocity distribution}

In the steady-state, one can integrate out the spatial coordinate $\mathbf{x}$ in the expression~\eqref{eq:potentialfree_solution} to obtain the marginal velocity distribution 
\begin{equation}
 \text{f}(\mathbf{v})=\int d\mathbf{v} \, p_s(\mathbf{x},\mathbf{v})\,.
\end{equation}
Unfortunately, a general analytical expression as in the case of the spatial density in Eq.~\eqref{eq:density} cannot be easily derived,
since the final result for $\text{f}(\mathbf{v})$ has a functional dependence on the shape of $u(\mathbf{x})$. 
 We argue that the 
spatial average $v_0$ of the swim velocity affects the shape of $\text{f}(\mathbf{v})$ in a trivial way, since $f$ can depend only on $\mathbf{v}/v_0$ as in the case $u(\mathbf{x})= v_0$. 
To proceed further, we restrict to the one-dimensional spatial modulation $u(x)$ of the swim velocity given by Eq.~\eqref{eq:swimvelocityprofile}.
With this assumption, the velocity distribution factorized between the different cartesian components of $\mathbf{v}$ so that, $\text{f}(\mathbf{v})=f_x(v_x) f_y(v_y)$. This allows us to study directly the velocity properties along a single component of the particle velocity namely, $v_x$. Below, we will drop the subscript $x$ for convenience of notation.

In Fig.~\ref{fig:marginal_prob_final}~(b), we show that the amplitude of the spatial variation of $u(x)$, determined by the parameter $\alpha$, plays a crucial role on the shape of the stationary velocity distribution $f(v)$.
While for $\alpha=0$ (spatially uniform case) $f(v)$ is given by a Gaussian distribution, as known in the literature~\cite{marconi2016velocity, fodor2016far, szamel2014self}, nonzero values of $\alpha$ induce a 
strong non-Gaussianity, 
reflected in particular by the heavy symmetric tails of the distribution (see the comparison between data and solid lines in Fig.~\ref{fig:marginal_prob_final}~(b)).
We remark that the non-Gaussian effects are controlled by the inhomogeneity of $u(x)$ alone, because $f(v)$ follows from the spatial integration of Eq.~\eqref{eq:potentialfree_solution}.
As $\alpha$ increases up to 1 (keeping $v_0$ fixed), the non-Gaussianity becomes more pronounced and the distribution more concentrated around $v=0$.
To quantify these effects, in Fig.~\ref{fig:marginal_prob_final}~(c), we study the velocity variance, $\langle v^2\rangle$, and the velocity kurtosis, $\langle v^4\rangle/\langle v^2\rangle^2$, as a function of $\alpha$.
We underline that, in this case, the two averages are performed by integrating out both $x$ and $v$.
As expected, 
$\langle v^2\rangle$ decreases with increasing $\alpha$ and vanishes in the limit $\alpha\to1$, where $f(v) \to \delta(v)$,
since the active particle remains stuck in the minima of $u(x)$.
Indeed, the larger $\alpha$, the longer is the time spent in the regions with the smallest velocity, namely $v \approx v_0 (1-\alpha)$, which is responsible for the lowering of $\langle v^2 \rangle$.
In the limiting case $\alpha\to1$, the variance vanishes, as the particle is not able to leave the region with $u(x)=0$, which means that we have $f(v) \to \delta(v)$.
On the other side, the growth of $\alpha$ induces the increase of the kurtosis from $3$, i.e. the value of the Gaussian distribution for $\alpha=0$, to higher values, eventually diverging in the limit $\alpha \to 1$.

Although we do not know the analytical form of $f(v)$ for $0<\alpha<1$, we can derive an expression for $\langle v^2\rangle$ by substituting the result for $p_s(x,v)$, given by the one-dimensional version of Eq.~\eqref{eq:potentialfree_solution}, and first calculating the integral over $v$:
\begin{equation}
\begin{aligned}
\label{eq:variance_v}
\langle v^2 \rangle& =  \int dx\,  \rho(x) 
\,u^2(x)= \int_0^L dx \mathcal{N} u(x)\\
&= L \mathcal{N}v_0 =v_0^2 \sqrt{1-\alpha^2} \,.
\end{aligned}
\end{equation}
In the second line, we have used the explicit form of $u(x)$ and the corresponding $\mathcal{N}$.
A similar strategy, allows us to predict the profile of the kurtosis: 
\begin{equation}
\label{eq:kurtosis_v}
\frac{\langle v^4 \rangle}{\langle v^2 \rangle^2} = \frac{\int dx \, u^3(x)}{\mathcal{N}\left(\int dx\, u(x)\right)^2}= \frac{3}{\sqrt{1-\alpha^2}}\left( 1+ \frac{3}{2} \alpha^2 \right)  \,.
\end{equation}
These exact predictions 
are shown in Fig.~\ref{fig:marginal_prob_final}~(c).
From Eq.~\eqref{eq:variance_v}, the variance of the reduced velocity distribution monotonically decreases to zero for $\alpha\to1$.
Likewise, the divergence of the kurtosis can directly be inferred from the denominator in Eq.~\eqref{eq:kurtosis_v}.



\section{Time-dependent properties}\label{sec:timedependent}

In this section, we focus on the steady-state temporal properties of the system, such as the velocity autocorrelation function, the mean-square displacement and, finally, the long-time diffusion.
To consider the diffusive properties of the system, 
In particular, we numerically study Eq.~\eqref{eq:AOUP} 
with the swim velocity landscape given by Eq.~\eqref{eq:swimvelocityprofile}, i.e. a swim velocity profile varying along one coordinate only. This choice allows us to restrict the numerical study to a single-coordinate only as in Sec.~\ref{sec:steadystate}.
The dynamics of each spatial coordinate is integrated over an infinite line, i.e. without periodic boundary conditions, in such a way that the periodic and bounded shape of $u(x)$ allows the system to perform standard diffusion for long times.

\subsection{Velocity autocorrelation function}

\begin{figure*}[t!]
\includegraphics[width=1\textwidth]{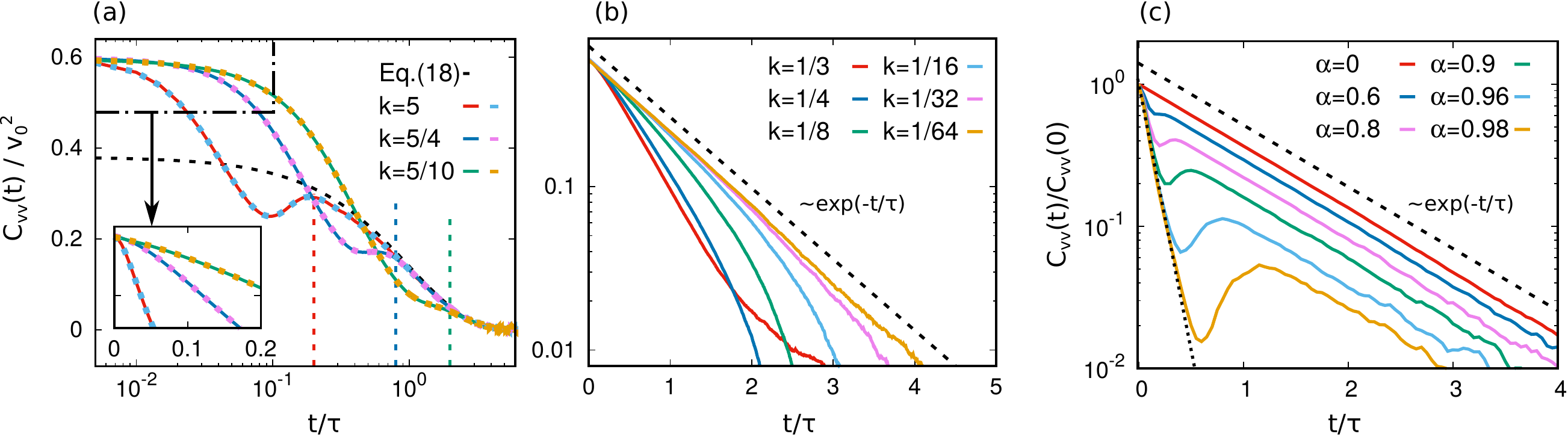}
\caption{Velocity autocorrelation function, $C_{vv}(t)=\langle v(t) v(0)\rangle$, as a function of the rescaled time, $t/\tau$.
Panels~(a) and~(b): $C_{vv}(t)/v_0^2$ by keeping fixed $\alpha=4/5$ for different values of $k$. In addition, in panel~(a), each $k=v_0\tau/S$ value is obtained by varying the parameters in two different ways, in particular, the dashed lines are obtaining by doubling both $v_0$, $\tau$ and by taking $S\to 4S$. 
The black dashed curve in panel~(a) is obtained from the theoretical prediction~\eqref{eq:corr_longtime} (normalized by $v_0^2$), valid beyond the vertical dashed lines, indicating the characteristic time scale $\tau/k = S/(v_0)$ according to the color legend.
The inset in panel~(a) zooms on a small time-window to highlight the behavior for small time.
Panel~(c):  $C_{vv}(t)/C_{vv}(0)$ for different values of $\alpha$ keeping fixed $k=v_0\tau/S=5/2$. 
Finally, the black dashed lines in panels~(b) and~(c) are guides for the eyes, showing the exponential function, $\exp(-t/\tau)$, while the black dotted one in panel~(c) is obtained by fitting the exponential function, $a\,e^{-t/b}$ ($a$ and $b$ are two fitting parameters). 
}
\label{fig:vv(t)_final.eps}
\end{figure*}

The steady-state velocity autocorrelation function, $C_{vv}(t)=\langle v(t) v(0)\rangle$, plays a relevant role because $C_{vv}(t)$ is related to the mean-square displacement and the long-time diffusion via the Green-Kubo relations. 
Hereafter, $\langle \cdot \rangle$ again represents the average over both position and velocity.
The three panels of Fig.~\ref{fig:vv(t)_final.eps} show $C_{vv}(t)$ as a function of $t/\tau$ for different values of the model parameters.
While the natural time-scale is provided by the persistence time $\tau$, the amplitude of the $C_{vv}(t)$ is trivially proportional to $v_0^2$.
As already discussed, the dynamics is controlled by two non-dimensional parameters, namely the amplitude $\alpha$ (which affects also the steady-state properties of the system) and $k=v_0 \tau/ S$, i.e. the ratio between the average persistence length $v_0 \tau$ 
of an active particle with constant swim velocity and the spatial period $S$ of $v_0(x)$.
This scaling is verified 
in Fig.~\ref{fig:vv(t)_final.eps}~(a), where $C_{vv}(t)/v_0^2$ is plotted as a function of $t/\tau$ for different 
$k$, each obtained with two distinct combinations of values of the parameters $S$, $\tau$ and $v_0$.
%
We first aim to 
understand the role of the ratio $k$ between the typical length scales of the system by keeping the value of $\alpha$ fixed. Here, we distinguish between different regimes, large and small $k$, shown in Fig.~\ref{fig:vv(t)_final.eps}~(a) and Fig.~\ref{fig:vv(t)_final.eps}~(b), respectively.

For large $k$, i.e. when the persistence length is larger than $S$ and, thus, $u(x)$ varies fastly, we can distinguish between two time-regimes: i) for small $t/\tau$, the autocorrelation $C_{vv}(t)$ decays exponentially with a typical relaxation time which strongly depends on $k$ (see the inset of Fig.~\ref{fig:vv(t)_final.eps}~(a)).
The typical time (vertical lines) for which 
this first relaxation takes place increases as $k$ is decreased. Afterwards, 
a non-monotonic shape in the profile of $C_{vv}(t)$ is observed till $C_{vv}(t)$ approaches another exponential relaxation $\sim e^{-t/\tau}$.
This last relaxation is uniquely controlled by $t/\tau$ and resembles that of an active particle with the homogeneous swim velocity $v_0$.


Next, in Fig.~\ref{fig:vv(t)_final.eps}~(b), we study $C_{vv}(t)$ for smaller values of $k$ at fixed $\alpha$.
Between $k=1/3$ and $k=1/4$, the curvature of the profile changes sign and we only observe a single time-decay regime which is faster than an exponential. 
When $k$ is decreased further, the typical time of this relaxation grows and the shape of $C_{vv}(t)$ approaches to an exponential, $\sim e^{-t/\tau}$, approximatively when $k$ is smaller than $k<1/32$.

To get further analytical insight, we multiply 
the one-dimensional version of Eq.~\eqref{eq:dynamics_xv} by $v(0)$ and take 
the average over $x$ and $v$, obtaining an effective equation for $C_{vv}(t)$: 
\begin{equation}
\label{eq:ODEcorr_longtime}
 \gamma\tau\dot{C}_{vv}(t) =- \gamma C_{vv}(t) + \gamma\tau \left\langle \frac{v^2(t)}{u(x(t))}\frac{\partial}{\partial x} u(x(t)) v(0) \right\rangle \,.
\end{equation}
When the oscillations of $v_0(x)$ are very rapid, one can replace $\frac{\partial}{\partial x} u(x) \sim \sin(x\, 2\pi/S)$ by its average, which vanishes.
In this way, for $t \gg S/v_0$, one can easily obtain the profile 
\begin{equation}
\label{eq:corr_longtime}
C_{vv}(t)\approx \frac{D_L(\alpha)}{\tau} e^{-t/\tau} \,,
\end{equation}
where $D_L(\alpha) \propto v_0^2 \tau$ is 
a function of $\alpha$ which is constant in time with $D_L(\alpha=0) =v_0^2 \tau$. This function will later be specified and interpreted as the long-time diffusion.
As shown in Fig.~\ref{fig:vv(t)_final.eps}~(a), the profile~\eqref{eq:corr_longtime} fairly agrees with numerical simulations after a transient time which depends on the value of $k$.
%
Moreover, for $k \ll1$, the persistence length is much smaller than $S$ and the particle velocity relaxes before the spatial variation of the swim velocity affects the dynamics of the system
and we can again neglect $\frac{\partial}{\partial x} u(x)$ in Eq.~\eqref{eq:ODEcorr_longtime} to recover the approximation for $C_{vv}(t)$ given by Eq.~\eqref{eq:corr_longtime} (see the black dashed curve in Fig.~\ref{fig:vv(t)_final.eps}~(b)).
In other words, the system behaves as a passive system with space-dependent diffusivity equivalent to that studied in Ref.~\cite{Breoni2021spatial}: 
this is clear by considering Eq.~\eqref{eq:AOUP_eta} with $\dot{\eta}=0$ so that $\eta \approx \sqrt{2} \chi$.

From Eq.~\eqref{eq:corr_longtime}, we also infer that the amplitude of the swim velocity oscillations (controlled by $\alpha$) cannot affect the relaxation of $C_{vv}(t)$, even if it determines the steady-state variance of the distribution.
To evaluate in detail the role of $\alpha$ on the velocity relaxation, we study $C_{vv}(t)/C_{vv}(0)$ for several values of $\alpha$ in Fig.~\ref{fig:vv(t)_final.eps}~(c).
We fix $k=5/4$ to consider a regime of the persistence length large if compared with the spatial period of $u(x)$.
The two relaxation times, in the long-time regime and in the small-time one, do not depend on $\alpha$, in agreement with prediction~\eqref{eq:corr_longtime}.
However, $\alpha$ affects the survival time of the small time-regime, so that larger values of $\alpha$ (close to the maximal value 1) mean that $C_{vv}(t)$ reaches very small values already in the first time regime, approaching zero for $\alpha=1$.
As a matter of fact, the second regime almost coincides with that of homogeneous active particles so that $u(x)=v_0$ and $\alpha=0$, except for the numerical factor $D_L(\alpha)$.
This means that for large times the structure of the velocity landscape (through the parameters $k$ and $\alpha$) is almost irrelevant and the relaxation of the velocity is mainly determined by the persistence $\tau$. 
Nevertheless, the shape of $u(x)$ is still relevant because it affects the long-time diffusion $D_L(\alpha)$ of the system through $\alpha$.
The first decay regime, instead, accounts for the microscopic details of the velocity landscape and indeed shows an explicit dependence on both $k$ and $\alpha$.
This quantifies the relaxation of the velocity towards a single well of the energy landscape, i.e. near a minimum of $u(x)$, where the probability is large and the velocity variance is small.
The larger the value of $\alpha$ (which cannot exceed 1 by definition), the larger the time needed to escape from a well of the energy landscape and to recover a diffusive behavior till to the limiting case $\alpha=1$ where the particle remains stuck in the minimum of the potential barrier.
However, we remind that the result for $\alpha=1$ is not physical because, when the swim velocity $u(x)\to0$ locally, the effect of the viscous solvent can be no longer neglected and the particle, instead of remaining stuck, performs Brownian translational diffusion due to the thermal agitation which here has been neglected.

\subsection{Mean-square Displacement}

\begin{figure*}[t!]
\includegraphics[width=1\textwidth]{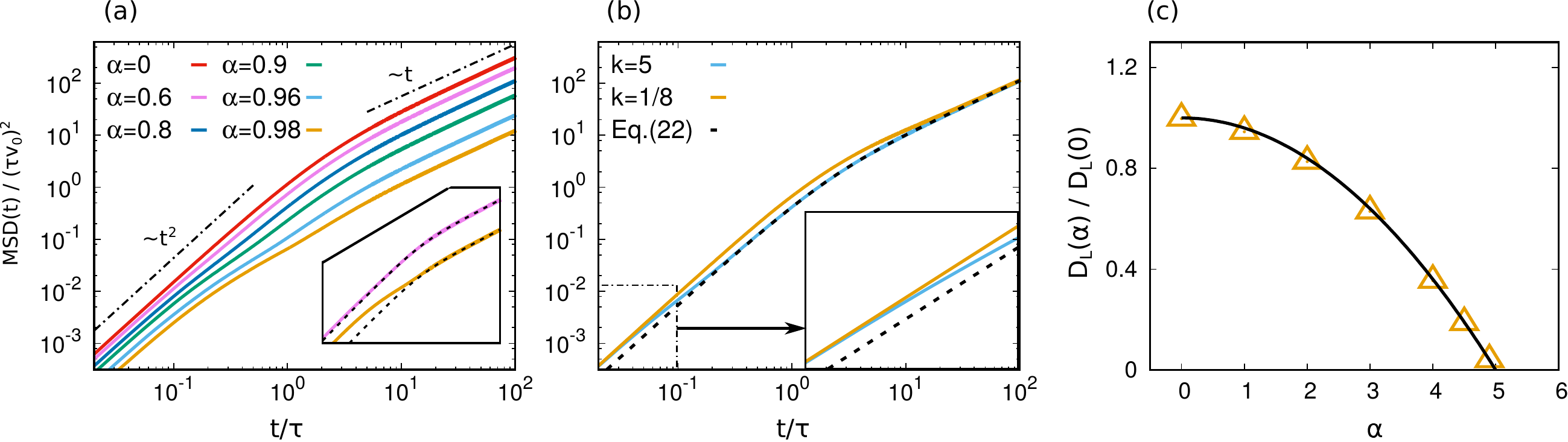}
\caption{Diffusive properties of the system. Panel~(a) and~(b): mean-square-displacement rescaled by $v_0 \tau$, defined as $\text{MSD}(t)=\langle (x(t)-x(0))^2\rangle$, as a function of the rescaled time, $t/\tau$. 
In panel~(a), we vary $\alpha$ keeping fixed $k=v_0 \tau/S=5/4$, while in the inset, we only focus on two curves to show the comparison with the theoretical predictions (black dashed lines), obtained by Eq.~\eqref{eq:MSDprediction}. Inset and main panel share the same axis.
Black dashed-dotted lines in the main panel are eye-guides for the ballistic ($\sim t^2$) and diffusive ($\sim t$) behavior. 
 in panel~(b), we vary $k$ keeping fixed $\alpha=0.9$, while, in the inset, we zoom on the region of the graph embodied in the dashed dotted rectangle. Again, black dashed lines plot Eq.~\eqref{eq:MSDprediction}. 
The insets of both panels~(a) and~(b) share the same legend as the corresponding main panels.
Panel~(c):  long-time-diffusion coefficient, $D_L(\alpha)/D_L(0)$ as a function of $\alpha$ (yellow triangles) and theoretical prediction~\eqref{eq:DLprediction} (solid black line). Here, the value of $k$ does not affect $D_L(\alpha)/D_L(0)$ in agreement with Eq.~\eqref{eq:DLprediction}.
}
\label{fig:MSD_final.eps}
\end{figure*}

Applying the Green-Kubo formula to the steady-state prediction~\eqref{eq:corr_longtime}, one can immediately obtain an analytic expression for the mean-square-displacement, defined as $\text{MSD}(t)=\langle (x(t)-x(0))^2\rangle$, which reads:
\begin{equation}
\begin{aligned}
\label{eq:MSDprediction}
\text{MSD}(t)&= \int^t_0 ds \int^s_0 ds' \langle v(s) v(s') \rangle \\
&\approx D_L(\alpha) \left[ t +\tau\left(e^{-t/\tau} -1\right)  \right] \,.
\end{aligned}
\end{equation}
The resulting approximation for the MSD has the same range of validity as Eq.~\eqref{eq:corr_longtime}, i.e. it holds for large times and small $k$. In addition, from Eq.~\eqref{eq:MSDprediction}, the constant $D_L(\alpha)$ can be easily interpreted as the long-time diffusion coefficient, which will be studied in detail as a function of the parameters of the model in the following subsection.

In Fig.~\ref{fig:MSD_final.eps}~(a), the $\text{MSD}(t)$ is numerically evaluated as a function of $t/\tau$ for several values of $\alpha$ and $k=5/4$.
The $\text{MSD}(t)$ displays a ballistic regime, i.e. $\propto t^2$, for small times, a crossing regime, so that $\propto t^{\beta}$ with $1<\beta<2$, for intermediate times, and, finally, a linear behavior $\propto t$ typical of standard diffusion.
As expected from the shape of the velocity correlations, the small-time regime of the $\text{MSD}(t)$ is only strongly affected by the value of $\alpha$ only if $\alpha\simeq1$. 
Moreover, as shown in Fig.~\ref{fig:MSD_final.eps}~(b) and in its inset, the $\text{MSD}(t)$ is almost insensitive to the value of $k$, upon appropriately rescaling the time as $t/\tau$ and the amplitude of the $\text{MSD}(t)$ with $ v_0^2\tau$ (i.e. the diffusion coefficient $D_L(0)$ of the active particle with constant $u(x)=v_0$). 
Finally, the diffusive (e.g. linear) regime of the $\text{MSD}(t)$ is purely determined by $\alpha$ and is independent of $k$, except for the trivial dependence on the scaling factor $v_0^2 \tau$. 

The analytic prediction~\eqref{eq:MSDprediction} reproduces the small-time regime only for small values of $\alpha$ while it fails for $\alpha \to 1$, as explicitly shown in the inset of Fig.~\ref{fig:MSD_final.eps}~(a).
In particular, the $t^2$-coefficient for $\alpha \to 1$ is much larger than $D_L(\alpha)/(2\tau)$, i.e. the $t^2$-coefficient predicted by Eq.~\eqref{eq:MSDprediction}.
The transient regime strongly depends on both $k$ and $\alpha$ and, in particular, occurs for a time that is larger as $k$ is decreased. 
As usual, the value of $\beta$ is not well-defined being a continuous curve interpolating from $\beta=1$ and $\beta=2$.
The long-time behavior is generally well-described by Eq.~\eqref{eq:MSDprediction}, as we detail in the following.

\subsection{Long time-diffusion coefficient}
From each curve of $\text{MSD}(t)$, we numerically extrapolate the long-time diffusion coefficient, $D_L(\alpha)$, by fitting the best linear function.  $D_L(\alpha)/D_L(0)$ is shown in Fig.~\ref{fig:MSD_final.eps}~(c) 
 as a function of $\alpha$ for a fixed value of $k$ that is irrelevant in the long-time regime as already shown in Fig.~\ref{fig:MSD_final.eps}~(b). 
Thereby, we verify that $D_L(\alpha)$ depends on $\tau$ and $v_0$ only via the global prefactor $v_0^2\tau$ which is nothing but the long-time diffusion $D_L(0)$ in 
the homogeneous case, 
while it decreases when $\alpha$ is increased.
This dependence has an intuitive physical interpretation: the larger the value of $\alpha$, the smaller is the minimal velocity ($\sim v_0(1 - \alpha)$) that the particle can assume during its motion.
Due to the increased probability $\rho(x)$ to find a particle in the regions with smaller swim velocity (cf. Eq.~\eqref{eq:density}), we expect that these regions 
dominate the diffusion properties, leading to a consequent decrease of the long-time diffusion coefficient with increasing $\alpha$.

To predict the shape of $D_L(\alpha)$, we resort to an argument similar to that already used in Ref.~\cite{caprini2020diffusion}, where the long-time-diffusion has been predicted for an active particle advected in a two-dimensional laminar flow.
At first, we calculate the effective persistence length $\ell$ of the active particle.
By separating the variables in the dynamics~\eqref{eq:AOUP_x} (with $D_t=0$) and integrating from $t=0$ to $t=\tau$, or equivalently from $x=0$ to $x=\ell$, we obtain 
\begin{equation}
\int_0^\ell \frac{dx}{u(x)} = \int_0^\tau dt \,\eta(t) \approx \tau \,,
\end{equation}
where, in the last approximation, we have used that the active force is roughly constant in the time-window given by the persistence time $\tau$.
Plugging in the shape of $u(x)$, performing the integral and accounting for the periodicity of $u(x)$, we get:
\begin{equation}
\ell(\alpha)  \approx \sqrt{1-\alpha^2} v_0 \tau\,.
\end{equation}
As expected, we explicitly find that $\ell(\alpha)$ decreases as $\alpha$ increases, in the same way as the $\langle v^2 \rangle$-result in Eq.~\eqref{eq:variance_v}. 
%
In analogy to the well-known relations in the case $u(x)=v_0$, the long-time diffusion coefficient can be estimated as the ratio between the square of the persistence length and the persistence time:
\begin{equation}
\label{eq:DLprediction}
D_L(\alpha) = \frac{\ell^2(\alpha)}{\tau} \approx v_0^2 \tau (1- \alpha^2) \,,
\end{equation}
This result reveals a fair agreement with data as shown in Fig.~\ref{fig:MSD_final.eps}~(c).
We mention that the prediction~\eqref{eq:DLprediction} has  already been put forward on the basis of symmetry arguments by Breoni et al.~\cite{Breoni2021spatial} in the case $\tau=0$, keeping fixed $D_L(0)=v_0^2\tau$. 
Our derivation applies also to that case because it holds for every value of $\tau$, included $\tau=0$ as a limiting case.

\section{Conclusion}

In this paper, we have studied the dynamical features of an active particle with a periodic spatial dependent swim velocity. 
The system is inspired by recent experiments obtained using bacteria that respond to external stimuli~\cite{arlt2018painting, frangipane2018dynamic} or Janus particles whose active force can be spatially modulated by a source of external light~\cite{buttinoni2012active, palacci2013living, lozano2016phototaxis}.
At first, we introduce a version of the active Ornstein-Uhlenbeck model suitably generalized to account for a spatial-dependent motility landscape.
Our AOUP allows us to recover a profile of the spatial density proportional to the inverse of the swim velocity, consistently with other active matter models. 
Describing the model in terms of particle position and effective velocity, we analytically find the exact steady-state solution of the Fokker-Planck equation: the density is peaked in the regions characterized by small swim velocity where the particles move slowly and the velocity distribution is Gaussian with a space-dependent kinetic temperature.
The reduced probability distribution of the velocity (averaging over the space) reveals a 
non-Gaussianity which becomes stronger as the amplitude of the swim velocity oscillations increases.
Then, we characterize the dynamics focusing on
 the velocity autocorrelation functions and mean-square displacement, which are numerically evaluated as a function of the parameters of the model.
We also provide an accurate theoretical prediction when the persistence length is larger than the period of the oscillation, i.e. when the swim velocity varies slowly in space. 
 The small-time regime of both observables is strongly affected by the details of the motility landscape, in agreement with Ref.~\cite{breoni2020active}.
Instead, in the large-time regime, the spatial variation of the swim velocity plays a less importnat role since the only relevant time-scale is the persistence time.
In this regime, we predict that the long-time diffusion coefficient decreases as the amplitude of the spatial oscillations of the the swim velocity increases.

A  final remark
concerns the experimental technique of "painting with bacteria" 
by Arlt et al.~\cite{arlt2018painting, arlt2019dynamics} and Frangipane et al.~\cite{frangipane2018dynamic}
mentioned in the introduction.  Our results suggest not only that the active agents would sit preferentially in the minima of the
active force $u(x)$, which can be tuned by light fields,  but also that the patterns so obtained are more stable in time against diffusion for large values of $\alpha$. In fact, the larger
the value of this parameter the smaller the diffusion.

In future studies, we will focus on the interplay between spatial dependent swim velocity and external confining force due to an external potential. In the framework of active colloids, the confining potential can be realized through acoustic~\cite{takatori2016acoustic} or optical traps~\cite{schmidt2018microscopic, schmidt2021non, aubret2021metamachines, ma2015enzyme}.
The introduction of a suitably generalized AOUP model~\eqref{eq:AOUP} for this situation paves the way to conveniently study problems of interacting particles with a spatial dependent swim velocity using 
generalizations of the Unified Colored Noise Approximation~\cite{jung1987dynamical, hanggi1995colored, maggi2015multidimensional,wittmann2017effective} 
or the Fox approach~\cite{fox1986,wittmann2017effective,sharma2017}.  
This opens up the possibility of obtaining interesting results concerning the probability distributions of position and velocity or analytical approximations for pressure and surface tension.
Next, it has been shown~\cite{lozano2016phototaxis, lozano2019propagating, jahanshahi2020realization} that Janus particles in optical landscapes experience a significant torque that aligns the particle along the intensity gradient. This torque needs to be included in future treatments of the problem.
Moreover, the effect of inertia should be considered for the case where the swim velocity depends on space~\cite{caprini2020inertial, nguyen2021active, lowen2020inertial}.
Finally, we mention a recent experimental and theoretical study by Sprenger et al., who proposed an experimental method to modulate the particle rotational diffusivity~\cite{sprenger2020active} in systems of active magnetic dumbbells.
The extension of our result to that case, i.e. by requiring that the persistence time has a spatial profile, represents a promising future perspective that could be addressed by employing similar theoretical methods.

\appendix

\section{Appendix A: Other version of AOUP with spatial dependent swim velocity}\label{app:AOUPMartin}

In this appendix, we will briefly review the AOUP dynamics proposed by Martin et al.~\cite{martin2020statistical} and will show that it differs from the AOUP model introduced in the present work.
This explains why we have obtained different results with respect to Ref.~\cite{martin2020statistical}, concerning for instance the steady-state density profile and the hydrodynamic approach.

Recently, Martin et al. have presented a generalization of the AOUP model to describe active particles with a space-dependent swim velocity, whose dynamics reads:
\begin{subequations}
\label{eq:AOUP_mod}
\begin{align}
\label{appeq:first}
&\dot{\mathbf{x}}= \sqrt{2 D_t}\boldsymbol{w}  +   \boldsymbol{\zeta} \\
\label{appeq:second}
&\tau(\mathbf{x}, t)\dot{\boldsymbol{\zeta}} = - \boldsymbol{\zeta} + D_a(\mathbf{x}, t)\sqrt{2}\boldsymbol{\chi}  \,.
\end{align}
\end{subequations}
In their approach, the term $\boldsymbol{\zeta}$ represents the active force, while $D_a(\mathbf{x}, t)>0$ is a spatial (time) dependent diffusivity and $\tau(\mathbf{x}, t)>0$ even includes the possibility that the persistence time depends on space.
To relate Eqs.~\eqref{eq:AOUP_mod} to the ABP dynamics~\eqref{eq:ABP}, one has to introduce the spatial-dependent swim velocity  which can be straightforwardly obtained from the variance of the Ornstein-Uhlenbeck process: 
\begin{equation}
u(\mathbf{x}, t)^2  = \frac{D_a(\mathbf{x}, t)}{\tau(\mathbf{x}, t)} \,,
\end{equation}
so that the following relation holds: 
\begin{equation}
\label{appeq:changeofvariables}
\boldsymbol{\zeta}_i = u(\mathbf{x}, t) \boldsymbol{\eta}_i \,.
\end{equation}
Now, we show that Eqs.~\eqref{eq:AOUP_mod} does not coincide with the dynamics~\eqref{eq:AOUP_x}, introduced in this paper.
By eliminating $D_a(\mathbf{x}, t)$ in favor of $u(\mathbf{x}, t)$ in Eq.~\eqref{appeq:second}, we obtain:
\begin{subequations}
\label{eq:AOUP_mod_v0}
\begin{align}
\label{eq:AOUP_mod_v0_first}
&\dot{\mathbf{x}}_i=\sqrt{2 D_t}\boldsymbol{w}_i  +  \boldsymbol{\zeta}_i \,, \\
\label{eq:AOUP_mod_v0_second}
&\tau \dot{\boldsymbol{\zeta}}_i = - \boldsymbol{\zeta}_i + u(\mathbf{x},t)\sqrt{2\tau}\boldsymbol{\chi}_i  \,,
\end{align}
\end{subequations}
upon assuming $\tau(\mathbf{x}, t)=\tau=\text{const}$ as in the case considered in the present paper.

Dynamics~\eqref{eq:AOUP_mod_v0} coincides with Eqs.~\eqref{eq:AOUP} only in the case $u(\mathbf{x}, t)=v_0$.
Indeed, only in that case, the change of variables provided by Eq.~\eqref{appeq:changeofvariables} can be obtained without including additional terms coming from the Jacobian of the change of variables.
In other words, if we plug expression~\eqref{appeq:changeofvariables} in Eqs.~\eqref{eq:AOUP_mod_v0}, the time-derivative on the left-hand-side of Eq.~\eqref{eq:AOUP_mod_v0_second} (in the static case, $u(\mathbf{x}, t)=u(\mathbf{x})$) reads:
\begin{equation}
\tau \dot{\boldsymbol{\zeta}}_i = \tau \frac{d}{dt} u(\mathbf{x}) \boldsymbol{\eta}_i = \tau \left(  u(\mathbf{x}) \frac{d}{dt}\boldsymbol{\eta}_i  + \boldsymbol{\eta}_i \dot{\mathbf{x}} \cdot \nabla u(\mathbf{x})\right) \,.
\end{equation}
The second term is not contained in Eq.~\eqref{eq:AOUP} and is responsible for the differences between our model and the one introduced by Martin et al.~\cite{martin2020statistical}.

\section*{Conflicts of interest}
There are no conflicts to declare.

\section*{Acknowledgements}
LC and UMBM warmly thank Andrea Puglisi for letting us use the computer facilities of his group and for discussions regarding some aspects of this research. 
LC and UMBM acknowledge support from the MIUR PRIN 2017 project 201798CZLJ. 
LC acknowledges support from the Alexander Von Humboldt foundation.
RW and HL acknowledge support by the Deutsche Forschungsgemeinschaft (DFG) through the SPP 2265, under grant numbers WI 5527/1-1 (RW) and LO 418/25-1 (HL).






\bibliography{SD}





\end{document}